\documentclass[sigconf,authorversion,nonacm]{acmart}
\usepackage{gnuplottex}
\usepackage{tikz}
\usepackage{multirow}
\usepackage{graphicx}
\usepackage{enumitem}
\usepackage{algorithm}
\usepackage{balance}
\usepackage[noend]{algpseudocode}
\usepackage{etoolbox}
\makeatletter
\patchcmd{\maketitle}{\@copyrightspace}{}{}{}

\makeatother
\renewcommand\footnotetextcopyrightpermission[1]{}

\begin{document}
\title{AppGNN: Approximation-Aware Functional Reverse Engineering using Graph Neural Networks}
\author{Tim~B{\"u}cher\textsuperscript{1}, Lilas~Alrahis\textsuperscript{2}, Guilherme~Paim\textsuperscript{3}, 
Sergio~Bampi\textsuperscript{3}, 
Ozgur~Sinanoglu\textsuperscript{2},~Hussam~Amrouch\textsuperscript{1}}
\affiliation{%
   \institution{\textsuperscript{1}Chair of Semiconductor Test and Reliability (STAR), University of Stuttgart, Germany\\
   \textsuperscript{2}Division of Engineering, New York University Abu Dhabi, UAE\\
   \textsuperscript{3}Informatics Institute, Universidade Federal do Rio Grande do Sul, Brazil \\ 
   E-mail: \textsuperscript{1}\{buecher, amrouch\}@iti.uni-stuttgart.de, \textsuperscript{2}lma387@nyu.edu }
   \city{}
   \state{}
   \country{}}
\newcommand{\red}[1]{\textcolor{red}{#1}}
\newcommand{\blue}[1]{\textcolor{blue}{#1}}
\definecolor{cadmiumgreen}{rgb}{0.0, 0.42, 0.24}
\renewcommand{\shortauthors}{B{\"u}cher and Alrahis, et al.}
\begin{abstract}
The globalization of the Integrated Circuit (IC) market is attracting an ever-growing number of partners, while remarkably lengthening the supply chain. Thereby, security concerns, such as those imposed by functional Reverse Engineering (RE), have become quintessential. RE leads to disclosure of confidential information to competitors, potentially enabling the theft of intellectual property.
Traditional functional RE methods analyze a given gate-level netlist through employing pattern matching towards reconstructing the underlying basic blocks, and hence, reverse engineer the circuit's function. 

In this work, we are the first to demonstrate that applying Approximate Computing (AxC) principles to circuits significantly  improves the resiliency against RE. This is attributed to the increased complexity in the underlying pattern-matching process.
The resiliency remains effective even for Graph Neural Networks (GNNs) that are presently one of the most powerful state-of-the-art techniques in functional RE. Using AxC, we demonstrate a substantial reduction in GNN average classification accuracy-- from $98\%$ to a mere $53\%$. 
To surmount the challenges introduced by AxC in RE, we propose the highly promising AppGNN platform, which enables GNNs (still being trained on exact circuits) to: (i)~perform accurate classifications, and (ii)~reverse engineer the circuit functionality, notwithstanding the applied approximation technique. AppGNN accomplishes this by implementing a novel graph-based node sampling approach that mimics generic approximation methodologies, requiring zero knowledge of the targeted approximation type. 

We perform an extensive evaluation targeting wide-ranging adder and multiplier circuits that are approximated using various AxC techniques, including state-of-the-art evolutionary-based approaches. We show that, using our method, we can improve the classification accuracy from $53\%$ to $81\%$ when classifying approximate adder circuits that have been generated using evolutionary algorithms, which our method is oblivious of.
Our AppGNN framework is publicly available under \url{https://github.com/ML-CAD/AppGNN}
\end{abstract}

\maketitle
\renewcommand{\headrulewidth}{0.0pt}
\thispagestyle{fancy}
\lhead{}
\rhead{}
\chead{This is the author's version of the work.
The definitive Version of Record is published in the 2022 International Conference On
Computer-Aided Design (ICCAD)}
\cfoot{}
\newlength{\textfloatsepsave} \setlength{\textfloatsepsave}{\textfloatsep} 
\begin{table}[!b]
\caption{Comparison of functional RE methods and their ability to identify exactly matching sub-circuits, sub-circuits with slight variations, and their applicability on AxC}
\begin{tabular}{cccc}
\hline
\textbf{Method} & \textbf{Exact matching} & \textbf{Variations} & \textbf{AxC} \\ \hline
Template matching & \multirow{2}{*}{\color{cadmiumgreen}{$\checkmark$}} & \multirow{2}{*}{\red{$\times$}} & \multirow{2}{*}{\red{$\times$}} \\ 
\cite{bsim:2014,wordrev:2013,patternmining:2012,circuitUnderstanding:2014} & & & \\\hline
Traditional ML~\cite{circuitrecognition:2019:DATE,circuitrecognition:2017:HOST} & \color{cadmiumgreen}{$\checkmark$} & \color{cadmiumgreen}{$\checkmark$} & \red{$\times$} \\ \hline
GNN-RE~\cite{GNNRE} & \color{cadmiumgreen}{$\checkmark$} & \color{cadmiumgreen}{$\checkmark$} & \red{$\times$} \\ \hline
\textbf{Proposed AppGNN} & \textbf{\color{cadmiumgreen}{$\checkmark$}} & \textbf{\color{cadmiumgreen}$\checkmark$} & \textbf{\color{cadmiumgreen}$\checkmark$} \\ \hline
\end{tabular}
\label{tab:comparison}
\end{table}

\section{Introduction}
Functional Reverse Engineering (RE) aims to analyze gate-level netlists that have been synthesized from a Register-Transfer Level (RTL) description of an Integrated Circuit (IC) design. A gate-level netlist contains information about the individual gates of a design and how they are interconnected to implement certain functionality. A description of which functionality is actually implemented however, is no longer present. It is the goal of functional RE to retrieve this information and reconstruct a high-level description of the functionality within the netlist~\cite{bsim:2014}. While the retrieval of a high-level description itself is challenging, it becomes even more of a task when analyzing approximate circuits.

Approximate Computing (AxC) achieves significant advantages over traditional computing concerning circuit area and power efficiency in scenarios where some loss of quality in the computed result can be tolerated~\cite{jie:ETS:2013}. For instance, in deep learning~\cite{zervakis:TC:2022, abreu:TCAS-I:2022} or media processing~\cite{paim:TCSVT:2020} such as audio, video, or image compression, where an exact computation is often not necessary due to the limited perception capability in humans~\cite{Venkatesan:ICCAD:2011}.

Despite the widespread employment of AxC, the research community did not investigate its security aspects so far. To the best of our knowledge, we are the first to evaluate the resilience of AxC to functional RE. In this work, we show that existing functional RE methods fail to reverse engineer approximate circuits. In the following we discuss the limitations of existing methods, which are also summarized in Table~\ref{tab:comparison}.

\vspace{-0.5em}
\subsection{Motivation and Research Challenges}
\begin{figure}[!t]
 \centering
 \includegraphics[width=0.8\linewidth]{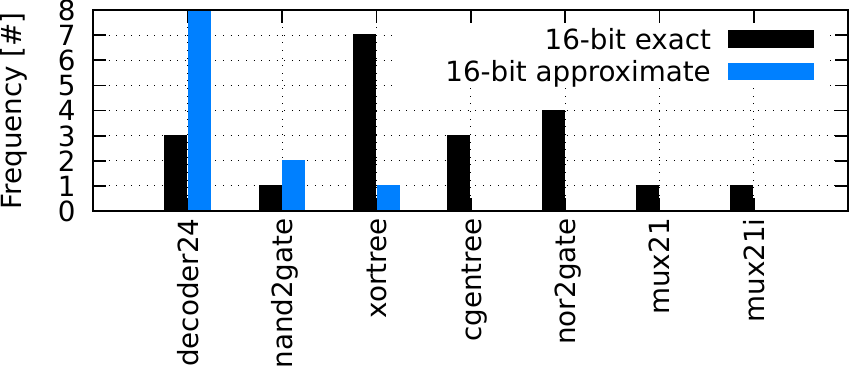}
 \caption{Number of identified sub-circuits by bsim~\cite{bsim:2014} in an exact 16-bit adder and an approximate 16-bit adder circuit.}
 \label{fig:bsim}
\end{figure}

\textbf{Golden Design Requirement:} Traditional functional RE methods that are based on template matching~\cite{10.1145/3193121,wordrev:2013,patternmining:2012,circuitUnderstanding:2014} have shown great potential in unveiling the functionality of gate-level netlists. However, these methods work by identifying sub-circuits and matching them to a \textit{"golden"} library of components known in prior. This makes them prone to erroneous classification caused by variations in the implementation of a module (i.e., structural or functional variations). With the rise of AxC, these methods finally fall short due to the vast design space that is opened up by the different possible methods of approximation. To investigate this, we employed the \textit{bsim} tool~\cite{bsim:2014}, one of the state-of-the-art algorithmic RE methods, on selected exact and approximate adder circuits. In our experiments, we have observed a mismatch between the types of recovered sub-circuits between the exact and the approximate implementations, i.e. they are not mapped to the same functionality, as Figure~\ref{fig:bsim} illustrates. \textit{This further highlights the fact that template matching-based approaches are not suitable for functional RE of AxCs.}

\begin{figure}[!t]
 \centering
 \includegraphics[width=0.9\linewidth]{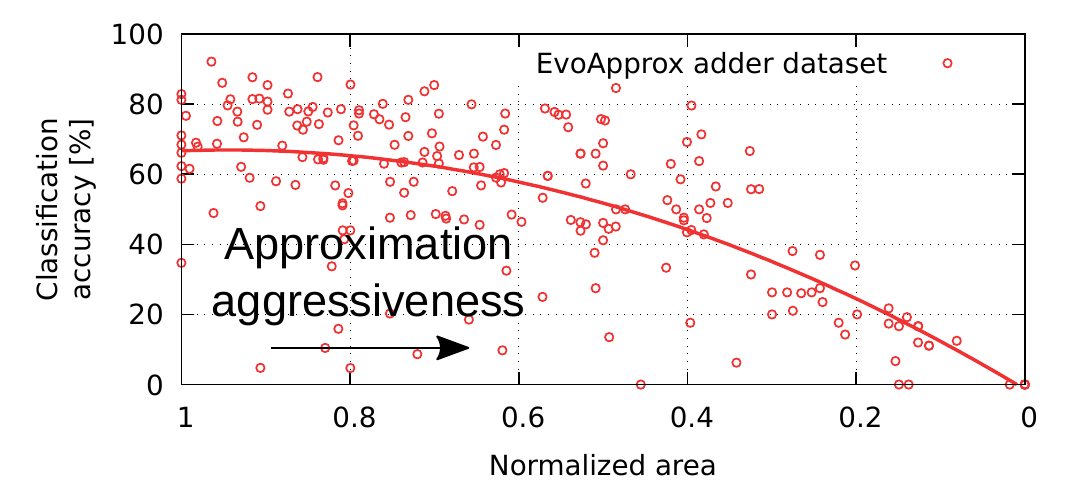}
 \caption{Node-level classification accuracy of GNN-RE~\cite{GNNRE} on the approximate adders from the EvoApprox~\cite{EvoApprox:2017} dataset. Approximation aggressiveness increases along the X-axis (drop in normalized area). GNN-RE classification accuracy drops with the increase in approximation aggressiveness.}
 \label{fig:gnn:performance}
\end{figure}

\textbf{Low Classification Performance:} Machine Learning (ML)-based approaches that employ Convolutional Neural Networks (CNNs)~\cite{circuitrecognition:2019:DATE,circuitrecognition:2017:HOST} as well as Graph Neural Networks (GNNs)~\cite{GNNRE} have been introduced for the task of functional RE~\cite{10.1145/3464959}. They promise to be more resilient against variations due to their generalization capabilities. \textit{However, their performance in the the presence of AxC has not yet been evaluated}. Therefore, we evaluated the state-of-the-art GNN-RE~\cite{GNNRE} for the task of classification of approximate adder circuits. While classification accuracy for exact adder circuits is on average $98\%$, once it was used for classification of approximate adder circuits, it reports an accuracy as low as $0\%$. Figure~\ref{fig:gnn:performance} further demonstrates this loss of performance and shows the relation between the aggressiveness of the approximation and the corresponding classification accuracy. The higher the aggressiveness in the approximation, the smaller individual circuits become and the more area is saved in the final integrated circuit~\cite{EvoApprox:2017}. At the same time, the loss in GNN classification accuracy becomes larger with increasing approximation aggressiveness.

\textbf{Research Challenges:} The discussion and experimental analysis above demonstrate that performing functional RE on approximate circuits is still an open research problem that imposes the following key research challenges. 
\begin{enumerate}
\item \textit{Handling inexact implementations:} Approximate circuits differ in their implementation compared to their exact circuit counterparts in terms of structure and functionality. A technique that is capable of generalizing to unseen approximated circuits (i.e., new structures) is required.
\item \textit{Handling different approximation types:} Different methods of approximation are possible for any given circuit. Handcrafted approximation methods can produce AxCs that structurally differ significantly from \textit{"black-box"} circuits that are generated using, for instance, evolutionary algorithms such as the EvoApprox~\cite{EvoApprox:2017} dataset (more details in Section~\ref{sec:background:AC}). Additionally, the aggressiveness of the approximation influences implementation details, further broadening the space that must be explored. \textit{Thus, a method that can generalize to different approximation types and levels is desired.}

\end{enumerate}
\vspace{-0.7em}
\subsection{Our Novel Concept and Contributions}
\begin{figure}[!t]
 \centering
 \includegraphics[width=\linewidth]{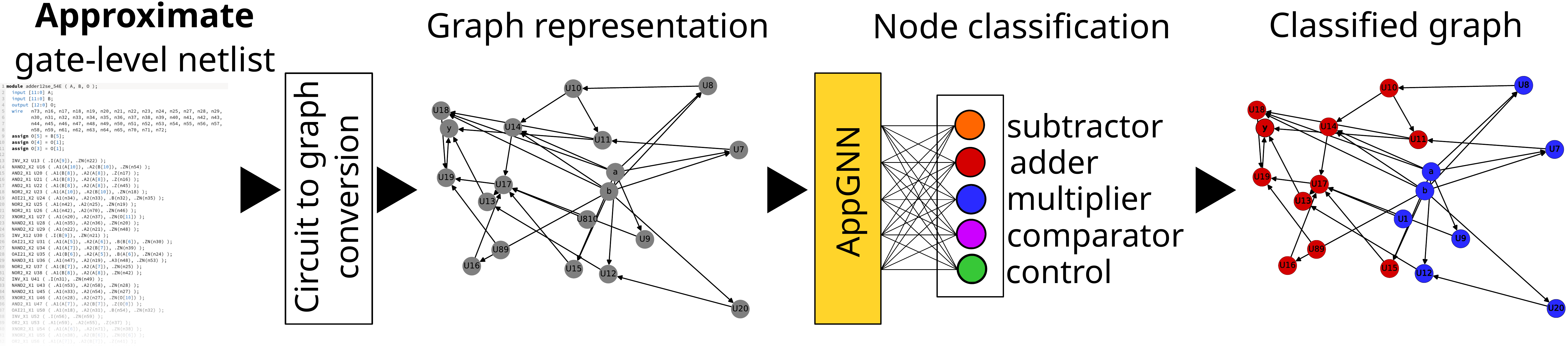}
 \caption{AppGNN approximation-aware node-level classification. Individual nodes in a given gate-level netlist are classified according to their role in a sub-circuit.}
 \vspace{-0.5em}
 \label{fig:motivational}
\end{figure}

To address the above challenges, we propose the \textit{AppGNN} platform that extends GNN-based functional RE to allow for the accurate classification of approximated circuits, all while training on exact circuits. Based on the fact that GNN-based RE relies on the structural properties of a circuit, we implement a novel graph-based sampling approach in AppGNN that can mimic a generic approximation of any design, in terms of structure, requiring zero knowledge of the targeted approximation type and its aggressiveness. Our platform operates on flattened (i.e., without hierachy) gate-level netlists and automatically identifies the boundaries between sub-circuits and classifies the sub-circuits based on their functionalities, whether the designs are approximate or not (see Figure~\ref{fig:motivational}). 

The novel contributions of this work are as follows: \\
(1)  A \textbf{comprehensive security analysis (Section~\ref{sec:analysis})} is provided. To the best of our knowledge, we are the first to investigate the resilience of AxC to functional RE.  \\
(2)  An \textbf{GNN-based functional RE platform (Section~\ref{sec:representation})} is developed. Our AppGNN approach does not rely on the exact matching of sub-circuits to a preexisting library, making it much more flexible in classifying AxCs.  \\
(3) A novel \textbf{graph-based sampling method (Section~\ref{sec:graph_sampling})} is proposed. Our AppGNN methodology mimics common adder approximation methods, on the graph-level, in order to improve the classification accuracy. \\
(4) {Our AppGNN framework is publicly available to the scientific community under: \url{https://github.com/ML-CAD/AppGNN}}

\textbf{Key results:} We perform an extensive experimental evaluation of AppGNN and GNN-RE~\cite{GNNRE} on various types of approximate circuits. We show that GNN-RE fails to reliably classify approximate circuits and that this failure is proportional to the aggressiveness of the approximation. To this end, we evaluate state-of-the-art GNN-RE and our AppGNN on the EvoApprox approximate adders and multiplier circuits~\cite{EvoApprox:2017}, as well various bit-width variations of Almost Correct Adder (ACA)~\cite{verma:DATE:2008}, Error-Tolerant Adder I (ETA-I)~\cite{zhu:TVLSI:2010}, Lower-part OR Adder (LOA)~\cite{mahdiani:TCAS-I:2010}, Lower-part Copy Adder (LCA)~\cite{gupta:TCAD:2013}, Lower-part Truncation Adder (LTA), Leading one Bit based Approximate (LoBA) multiplier~\cite{garg:JET:2020} and Rounding-based Approximate (RoBA) multiplier~\cite{reza:TVLSI:2017}, all of which are described in detail in Section~\ref{sec:background:AC}. Over all these extensive datasets, our approach achieves an improvement in classification of approximate adders and multipliers of up to $28$ percent points compared to GNN-RE, reaching a classification accuracy of up to $100\%$ for some circuits.
\vspace{-0.5em}
\section{Background and Related Work}
\label{sec:background}
In this section, we present a brief introduction to AxC along with the datasets used in this work. We will also explain the underlying principles of GNNs and functional RE.
\vspace{-0.6em}
\subsection{Approximate Computing (AxC)}
\label{sec:background:AC}

AxC is established as a new paradigm to boost design efficiency by trading the intrinsic error resiliency of several applications such as signal, image, video processing~\cite{paim:TCSVT:2020} and ML~\cite{zervakis:TC:2022, abreu:TCAS-I:2022}.
Driven by the high potential for energy savings, designing approximate functional units has attracted significant research interest~\cite{zervakis:ASPDAC:2021}.
In the following, we review approximate adder and multiplier functional units that we employ in this work as a case study. It is noteworthy that our work is not limited to certain type of approximate circuits. 
\vspace{-1em}
\subsubsection{Approximate Adders (AxAs)} AxAs can generally be classified as (i) Lower-Part Adders (LPA) with low-magnitude frequent errors, (ii) Block-based Speculative Adders (BSA), which result in high-magnitude infrequent errors, and (iii) Evolutionary Approximate (EvoApprox) circuits generated using evolutionary algorithms.

\noindent\textbf{Lower-Part Adders (LPAs).} LPAs split their operation into an exact part with $w-k$ Most Significant Bits (MSBs) and an approximate part with $k$ Least Significant Bits (LSBs), where $w$ is the bitwidth and $k$ is the approximation level. The higher $k$, the higher the errors in the LPA class. In this work, we consider the following LPAs: 
\begin{itemize}[noitemsep, topsep=0.1cm,leftmargin=0cm,itemindent=*] 
 \item LTA which truncates $k$ LSBs of the output with logic 0. 
 \item LCA~\cite{gupta:TCAD:2013} copies the $k$ LSBs of an operand to the $k$ output LSBs. 
 \item LOA~\cite{mahdiani:TCAS-I:2010} atributes a bitwise OR logic of the $k$ LSBs operands to the $k$ output LSBs.

\item ETA-I~\cite{zhu:TVLSI:2010} performs a carry-less sum with a bitwise XOR of the operands (i.e., a propagate operation).
To compensate for the missing carry, if a carry-generate (i.e., a bitwise AND of the operands) is true in a bit of the approximate part, the previous output LSBs are set to 1. 
An OR-logic chain propagates the set-to-one command from the generate position up to the LSB.

\end{itemize}

\noindent\textbf{Block-based Speculative Adders (BSA).} BSAs split their operation into blocks of $m$ width, where $m$ is related to the approximation level. 
The higher $m$ is, the lower are the errors and the benefits of the BSAs class.
In this paper, we consider the following BSA.

ACA~\cite{verma:DATE:2008} implements $m$-bit overlapping blocks to speculate the carry operation. ACA generates exact output values for the first $m$ LSBs. Each output bit at the position $m$ up to the MSB is independently generated by overlapping adder blocks that speculate $m$ bits of the operands. 

\vspace{-0.8em}
\subsubsection{Approximate Multipliers (AxMs)} AxMs are commonly designed by (i) mathematically refactoring the multiplication for eliminating parts of its equation and (ii) detecting the Leading One Bit (LOB) position of the operands for discarding the hardware that computes unnecessary leading zeros and (iii) truncating part of the LSBs. 
We employ the following multipliers that comprehend these strategies as case studies in this work.
 
\begin{itemize} [noitemsep, topsep=0.1cm,leftmargin=0cm,itemindent=*] 
 \item RoBA multiplier~\cite{reza:TVLSI:2017} simplifies the multiplication process by using numbers equal two to a power $n$ ($2^n$). The multiplication is factored as $A\times B = (A_r - A) (B_r -B) + A_r B + A B_r - A_r B_r $. Then, the $A_r B $, $A_rB_r $ and $A B_r $ sub-expressions are approximated by simple shift operations. Thus, the multiplication is simplified for $A \times B\approx A_r B + A B_r - A_r B_r$. A sweep process in the operands bits finds the nearest value of $A_r$ and $B_r$. The $A_{r}[i]$ is logic 1 in two cases. In the first case, $A[i]$ is set to logic 1, and all the bits on its left side are logic 0 while $A[i-1]$ is logic 0. In the second case, when $A[i]$ and all its left-side bits are logic 0, $A[i-1]$ and $A[i-2]$ are both logic 1. The circuit employs $B_r$ by the same process of $A_r$.
 
 \item LoBA multiplier~\cite{garg:JET:2020} splits the multiplication in smaller fixed-width multiplier blocks. LoBA multiplier identifies the LOB position and defines two $k$-bit new operands $A_{kh}$ and $B_{kh}$, which are multiplied and shifted left based on the LOB position. It multiplies the higher k-bit part of the operands starting from the LOB position suppressing leading zeros and truncating the lower part operands. Therefore, LoBA reduces the length of the multiplication to just $k$ bits. We refer as LoBA the LoBA0 configuration in~\cite{garg:JET:2020}.

\end{itemize}

\vspace{0.5em}
\subsubsection{Evolutionary-based Approximate (EvoApprox)}
EvoApprox adder and multiplier circuits are automatically generated by a genetic algorithm, a method inspired by natural selection that mimics biological evolution~\cite{EvoApprox:2017}. 
EvoApprox circuits are created by an extensive design exploration which evolves strong candidates and discards weak generations of circuits with a given error metric. We employ the EvoApprox library made available online by~\cite{EvoApprox:2017}.

\begin{figure*}[!t]
\centering
 \includegraphics[width=0.9\textwidth]{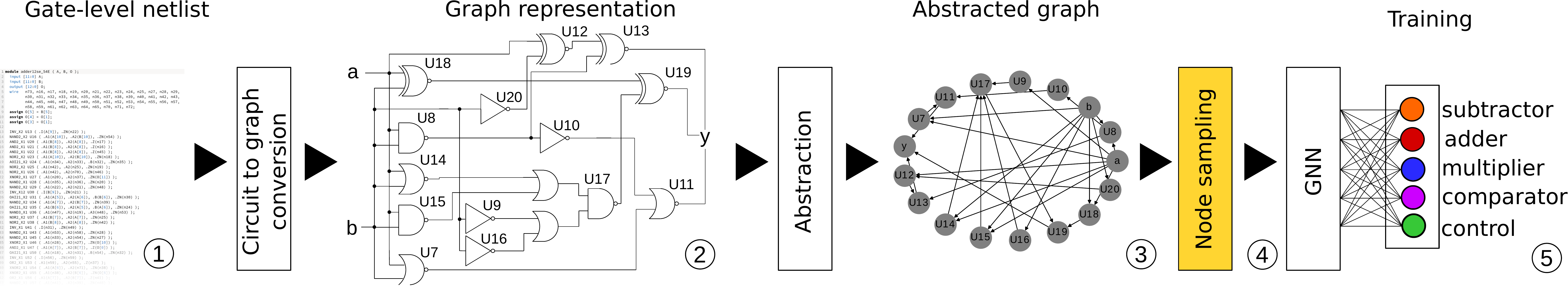}
 \caption{AppGNN work flow. Exact gate-level netlists are transformed into a graph representation. Node sampling is performed on the graph-level before the sampled graph is added to the training dataset and for the GNN to perform training.}
 \label{fig:flow}
\end{figure*}
\vspace{-3mm}
\subsection{Graph Neural Networks (GNNs)}
\label{sec:GNN}
GNNs perform graph-representation learning on graph-structured data and generate a vector representation (i.e., embedding) for each node in a given graph to be used for a desired task such as node classification. The embeddings are generated based on the input features of the nodes and the underlying graph structure so that similar nodes in the graph are close in the embedding space.

Let $G(V, E)$ denote a graph, where $V$ represents its set of nodes, and $E$ represents its set of edges. Each node in the graph $v \in V$ is initialized with a feature vector $x_v$ that captures its properties. In GNNs, typically an $Aggregate$ function collects the node information from the direct neighborhood of node $v$, denoted as $N(v)$. An $Update$ function updates the current embedding of node $v$ by combining its previous state with the aggregated information. GNNs run such a \textit{neighborhood aggregation} procedure for $L$ rounds (increasing the depth of the aggregated information)~\cite{kipf2016semi}.

GNNs mainly differ based on the $Aggregate$ and $Update$ functions used. The fundamental Graph Attention Network (GAT)~\cite{velivckovic2017graph} measures the importance (weight) of the edges during the aggregation phase. It employs a multi-head attention mechanism of $K$, in which the layer $l-1$'s information propagates to layer $l$ as follows;
\begin{equation}
{h}_{v}^{l}=
\left|\begin{matrix}
\\ 
\\ 
\end{matrix}\right|_{k=1}^K \sigma\begin{pmatrix}
\sum_{u\in N(v)} \alpha_{u,v}^k {W}^k {h}_{v}^{l-1}

\end{pmatrix}
\\
\end{equation}
\begin{equation}
 \alpha_{u,v}^k={LeakyReLU}
\begin{pmatrix}
({a}^k)^\intercal[{W}^k{h}_u \parallel {W}^k {h}_v]
\end{pmatrix}
\end{equation}
Where $h_{v}^{l}$ denotes the generated embedding of node $v$ at the $l^{th}$ round, $h_{v}^{0}={x}_{v}$, and $\sigma(.)$ is a non-linear activation function (e.g., \texttt{ReLU}). $\alpha_{u,v}$ specifies the weighting factor (i.e., importance) of node $u$'s features for node $v$, which is computed as a byproduct of an attention mechanism ${a}$. The multi-head attention mechanism replicates the aggregation layers $K$ times, each replica having different trainable parameters ${W}^{k}$, and the outputs are feature-wise aggregated using a concatenation operation as described in Equation~(1). The final embedding vector of node $v$, after $L$ layers, is as follows:
\begin{equation}
{z}_{v}= {h}_{v}^L
\end{equation}

\subsection{Functional Reverse Engineering (RE)}
Without access to the RTL description of a design, functional RE is one option to gain further detailed insight on the functions that are implemented inside a circuit.
However, gate-level netlists are inherently difficult to analyze, since structural information about the function and boundaries of sub-circuits is usually omitted during the synthesis process. This leaves only the high-level circuit description and list of gates and their interconnections to be analyzed. Therefore, automated algorithmic analysis tools, such as~\cite{bsim:2014}, have been introduced. Their functionality is based on partitioning the gate-level netlist by identifying replicated bitslices and, in turn, aggregating them into individual candidate modules. Afterwards, using formal verification methods, these candidate modules are matched against a \textit{"golden"} component library containing reference circuits in order to infer their high-level functionality, e.g. adder, multiplier, subtractor, etc~\cite{bsim:2014,wordrev:2013,patternmining:2012,circuitUnderstanding:2014}.

A shortcoming of pattern matching approaches like these is that they are extremely dependant on the quality and size of the \textit{"golden"} component library that is used. Due to this, components that differ from the "standard" implementation, for instance of an adder circuit, will not be classified correctly and may remain undetected~\cite{GNNRE}. Additionally, formal verification methods can be very resource demanding in terms of compute power, limiting their applicability. Therefore, other methods have been proposed in literature.

Recently, ML-based methods have shown great potential~\cite{GNNRE}, achieving high classification accuracy at only a one-time training effort. In GNN-RE, the gate-level netlist of a circuit is first transformed into a graph representation. The graph representation preserves the structure of the netlist, captures the features of each gate (for instance gate type, number of inputs/outputs, etc.) and also encodes the neighborhood of each gate. This way, the graph representation encodes both the structural and functional attributes of each node and it's surrounding circuitry. Then, GNN-RE employs a GNN to learn on the graph representation of the circuit to predict the sub-circuit each gate belongs to (i.e., node classification task).

In any GNN implementation, the quality of the model depends on the quality of the training data set that is used as a ground truth. Including a wide range of different implementations and bit-widths for each high-level component (e.g. adder, multiplier, etc.), the ML model can learn to generalize and become robust to variations.

Functional RE can be very helpful to various entities such as IC designers to verify to correctness of a chip. For instance, using functional RE, Hardware Trojans (HTs) or Intellectual Property (IP) infringements in competitor products can be detected~\cite{circuitrecognition:2017:HOST,circuitrecognition:2019:ASPDAC,circuitrecognition:2019:DATE}. On the flipside, functional RE can potentially also be abused to gain insights into patented designs and steal IP~\cite{REofEmbeddedSystems:2011}.
\newcommand{\bettercircle}[1]{\raisebox{.5pt}{\textcircled{\raisebox{-.9pt} {#1}}}}

\vspace{-0.5em}
\section{Our Proposed GNN-based Reverse Engineering: AppGNN}

\begin{figure}[!t]
 \centering
 \includegraphics[width=0.65\linewidth]{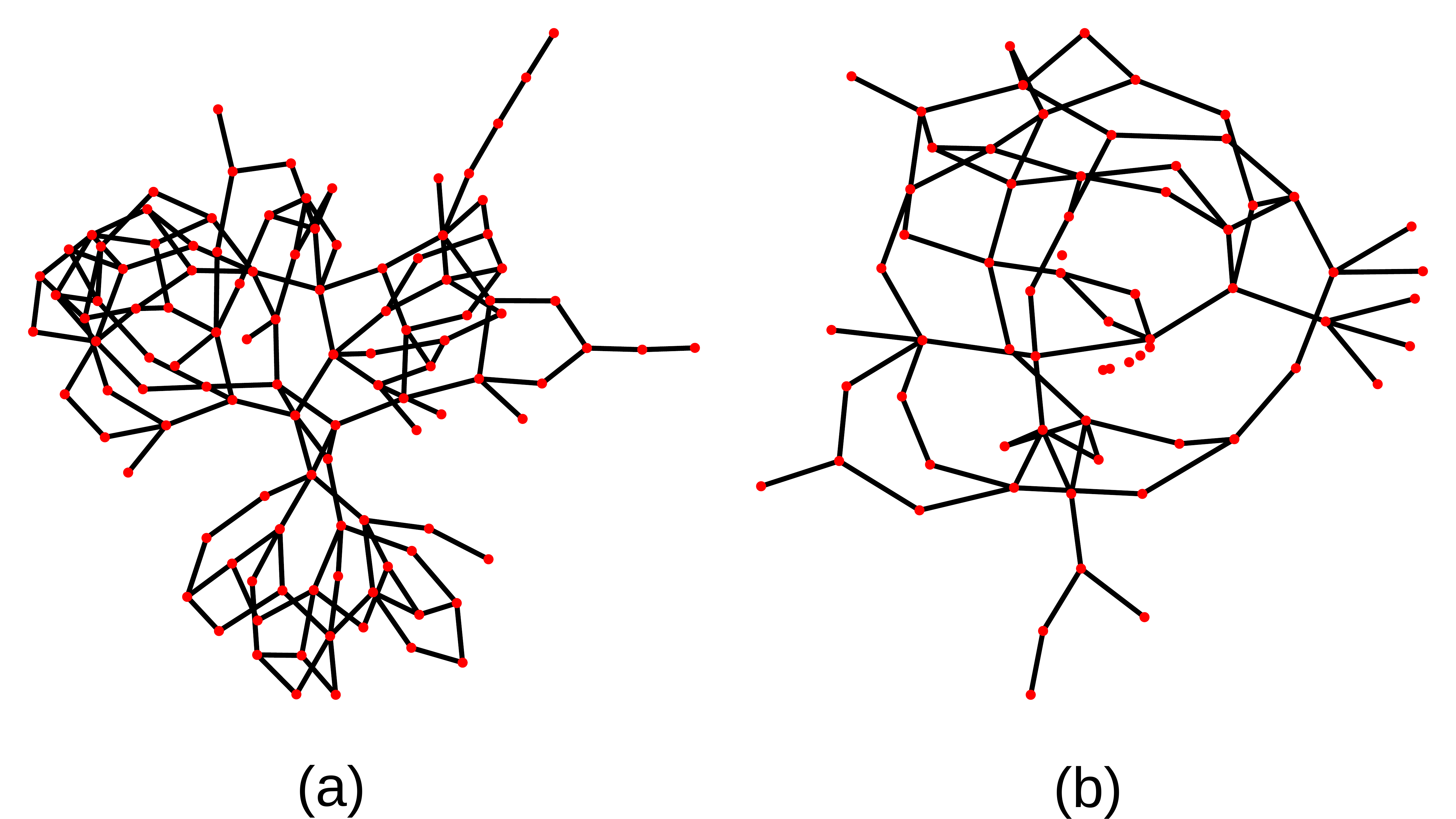}
 \caption{The graph representation of two 12-bit adder circuits. (a) shows the graph of an exact 12-bit adder circuit, (b) displays a LCA-based 12-bit AxC adder circuit. The structure and the number of gates differ between the two versions.}
 \label{fig:approximation:3bit}
\end{figure}
In the following, we describe our concept of AppGNN. We first cover the assumptions that we make. Then, we explain the graph representation for our GNN model, the datasets that we use for training and finally demonstrate our node sampling techniques.

The overall work flow of AppGNN is illustrated in Figure~\ref{fig:flow}.
We start in step \bettercircle{1} with a gate-level netlist, which is transformed into a graph (Section~\ref{sec:representation}) and abstracted in steps \bettercircle{2} and \bettercircle{3}. In step \bettercircle{4}, we perform node sampling using either random node sampling (Section~\ref{sec:random:sampling}) or leaf node sampling (Section~\ref{sec:leaf:sampling}). In final step \bettercircle{5}, the sampled graph is added to the training dataset (Section~\ref{sec:dataset}).
\vspace{-0.5em}
\subsection{Threat Model and Assumptions}
We perform functional RE on AxCs under the following assumptions, which are consistent with prior work~\cite{bsim:2014,wordrev:2013,patternmining:2012,circuitUnderstanding:2014,circuitrecognition:2017:HOST,circuitrecognition:2019:ASPDAC,circuitrecognition:2019:DATE,GNNRE}. We assume that the gate-level netlist of the design that is being analyzed has been correctly retrieved, either by deriving it from the physical chip~\cite{stateOfTheArtRE:2009} or by other means (e.g., access to layout information). In particular, access to the RTL source code of the design is not available. We make no assumptions about the given netlists that are analyzed. In particular, we do not assume any knowledge about the used approximation technique or the aggressiveness of the approximation. In fact, we do not assume it to be an approximate circuit at all. \textit{This allows us to develop a generic approach and to operate on exact circuits as well as approximate ones}.
\vspace{-0.5em}
\subsection{Why do Approximate Circuits Appear to be Reverse-Engineering Resilient?}
\label{sec:analysis}
Approximate circuits can differ substantially from their exact counterparts in terms of their general graph structure, gate count and gate type. As an example, Figure~\ref{fig:approximation:3bit} illustrates the graph representation of two 12-bit adder circuits. In the figure, (a) displays the connectivity of the individual gates of an exact adder circuit, (b) displays this for an AxC adder that has been generated using LCA~\cite{gupta:TCAD:2013}. Six of the twelve primary outputs are approximated (copied directly from the input to the output). As the figure illustrates, the graph structure of both circuits and their node count is dissimilar. 

\begin{figure}[!t]
 \centering
 \includegraphics[width=0.8\linewidth]{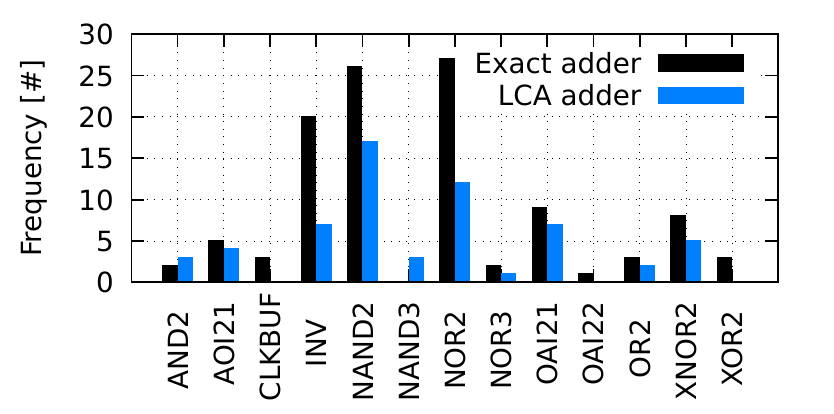}
 \caption{Histogram of the occurance of individual gate types in a 12-bit exact adder implementation and a LCA-based 12-bit AxC adder circuit.}
 \label{fig:hist:gatetypes}
 \vspace{2mm}
\end{figure}

\begin{table}[!t]
\caption{The datasets used in the training of AppGNN}
\resizebox{0.45\textwidth}{!}{%
\begin{tabular}{cccc}
\hline
\textbf{Datasets} & \textbf{\#Nodes} & \textbf{\#Circuits} & \textbf{Source} \\ \hline
Add-Mul-Mix & 15,582 & 7 & \multirow{7}{*}{GNN-RE~\cite{GNNRE}}\\
Add-Mul-Mux & 21,602 & 6 & \\ 
Add-Mul-Combine & 14,288 & 6 & \\ 
Add-Mul-All & 51,472 & 19 & \\ 
Add-Mul-Comp & 15,898 & 6 & \\ 
Add-Mul-Sub & 18,206 & 6 & \\
Add-Mul-Comp-Sub & 19,151 & 6 & \\ \hline
Adder 8bit & 469 & 8 & \multirow{4}{*}{This work (\textbf{AppGNN})}\\
Adder 9bit & 792 & 9 & \\
Adder 12bit & 981 & 9 & \\
Adder 16bit & 1314 & 9 & \\ \hline
\label{tab:training:dataset}
\end{tabular}
}
\vspace{-3mm}
\end{table}

In addition to the graph structure and gate count being different, the type of used gates in both circuits varies as well, as Figure~\ref{fig:hist:gatetypes} shows. It can be seen from the figure that some gate types appear exclusively in one of either circuits, such as \texttt{CLKBUF}, \texttt{OAI22} (Or-And-Invert) and \texttt{XOR2} which only appear in the exact implementation. Conversely, \texttt{NAND3} is exclusively used in the approximate implementation. The remaining shared gate types are used with different frequencies in the two implementations. For example \texttt{INV} gates are used $2.8$ times (20 vs. 7) and \texttt{NOR2} gates are used $2.25$ times (27 vs. 12) more frequently in the exact circuit compared to the approximate one. However, some similarity remains. Gate types \texttt{AND2}, \texttt{AOI2} (And-Or-Invert), \texttt{NOR3}, \texttt{OR2}, \texttt{OAI21} and \texttt{XNOR2} appear with similar frequency in the two implementations.

\begin{figure}[!t]
 \centering
 \includegraphics[width=0.8\linewidth]{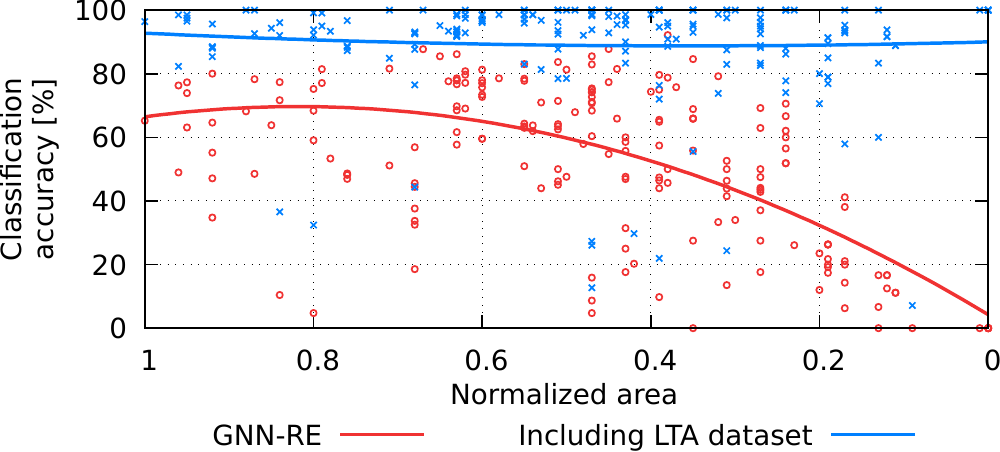}
 \caption{Node-level classification accuracy on the EvoApprox~\cite{EvoApprox:2017} adders dataset of GNN-RE and when the LTA dataset of approximate adders is included in the training.}
 \label{fig:gnn-re-lta}
 \vspace{-1mm}
\end{figure}

Our experiments show that when we extend the original GNN-RE dataset of exact circuits, as displayed in Table~\ref{tab:training:dataset}, by another set of approximate circuits, the classification accuracy on approximate circuits as a whole can be improved. Specifically, we add the dataset of LTA adders (see Table~\ref{tab:approx:dataset}) to the training dataset. The resulting improvement in classification accuracy on the EvoApprox adders can be seen in Figure~\ref{fig:gnn-re-lta}. The average node-level classification accuracy on the EvoApprox adders dataset increased from $53.8$\% (original GNN-RE) to $89.6$\% (GNN-RE extended by the LTA dataset). 

The above analysis shows that once the GNN is exposed to approximated structures during training, the classification accuracy on AxCs can be improved. However, this setup will require the user to access some AxC scripts to generate the required dataset, which may not be available. The analysis motivates us to extend the GNN training stage itself to manage AxC while still training on exact circuits. In order to improve the classification accuracy of a GNN regardless of the dataset that is added to the training process, we propose a node sampling approach which aims to mimic a generic approximation technique. Next, we discuss the main steps of our AppGNN framework, starting with circuit-to-graph conversion.

\begin{table}[!t]
\caption{Adders and multipliers in our evaluation dataset}
\resizebox{0.4\textwidth}{!}{%
\begin{tabular}{cccc}
\hline
\textbf{Evaluation dataset} & \textbf{Circuit type} & \textbf{\#Nodes} & \textbf{\#Circuits} \\ \hline
ACA~\cite{verma:DATE:2008} & adder & 2618 & 28\\ \hline
ETA-I~\cite{zhu:TVLSI:2010} & adder & 2283 & 28 \\ \hline
LOA~\cite{mahdiani:TCAS-I:2010} & adder & 2210 & 28 \\ \hline
LCA~\cite{gupta:TCAD:2013} & adder & 2282 & 28 \\ \hline
LTA & adder & 2028 & 28 \\ \hline
EvoApprox-Add~\cite{EvoApprox:2017} & adder & 14901 & 214 \\ \hline 
LoBA~\cite{garg:JET:2020} & multiplier & 43694 & 31 \\ \hline
RoBA~\cite{reza:TVLSI:2017} & multiplier & 31077 & 5 \\ \hline
EvoApprox-Mul~\cite{EvoApprox:2017} & multiplier & 78155 & 159 \\ \hline
Exact circuit & adder & 467 & 5 \\ \hline
Exact circuit & multiplier & 13701 & 6\\ \hline
\label{tab:evaluation:dataset}
\end{tabular}
}
\label{tab:approx:dataset}
\vspace{-6mm}
\end{table}

\vspace{-0.5em}
\subsection{Graph Representation}
\label{sec:representation}
The flat netlist is first transformed into a directed graph representation, which retains a notion of order to be able to reason about successor and predecessor nodes. In the graph, each gate in the netlist corresponds to a specific node. Edges represent connections (wires) between individual gates/nodes. This transformation from a netlist to a graph will retain all structural information present in the netlist, e.g. the connectivity between individual gates will be represented using an adjacency matrix. In order to retain information about the individual gates, such as the gate type (e.g. \texttt{XOR}), whether gates are Primary Inputs (PI) or Primary Outputs (PO) or the input and output degree, each node will keep a reference to a feature vector $x$ with length $k$ describing these details. The length of $k$ is largely determined by the number of available gate types in the used technology library\footnote{In our experiments, a 14nm FinFET library containing 24 individual gates was used.}, in our case $k=24$.

\begin{figure}
 \centering
 \includegraphics[width=0.75\linewidth]{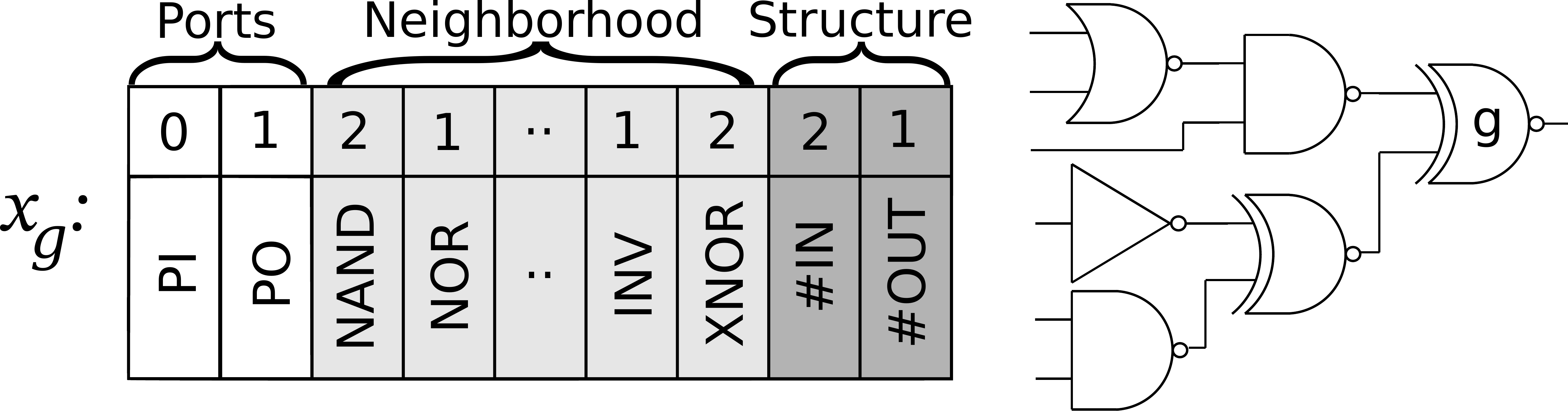}
 \caption{Feature vector $x_g$ of node g, adapted from~\cite{GNNRE}.}
 \label{fig:featurevector}
\end{figure}

Figure~\ref{fig:featurevector} shows an example of a netlist that contains various gate types. In this example, node $g$ is a primary output (since it is a leaf node) which is captured in the first two fields of the feature vector $x_g$ (\textit{Ports} section in Figure~\ref{fig:featurevector}). The feature vector also contains information about the gate type of $g$ (\texttt{XNOR}) as well as the gate types and number of occurrences of other gates present in the two-hop neighborhood\footnote{The two-hop neighborhood of a node $g$ includes all nodes directly adjacent to $g$ and all nodes that in turn are adjacent to these nodes.} of $g$. In this example, the 2-hop neighborhood contains two \texttt{NAND2}, two \texttt{XNOR}, one \texttt{NOR} and one \texttt{INV} gate. This neighborhood information is stored in the following fields of $x_g$ (\textit{Neighborhood} section in Figure~\ref{fig:featurevector}). The last two fields (\textit{Structure} section in Figure~\ref{fig:featurevector}) capture the input (2) and output (1) degree of $g$.

A feature matrix $X\in\mathbb{R}^{n\times k}$, with $n$ describing the total number of nodes, will aggregate the feature vectors of all nodes. The feature matrix $X$ is then standardized by removing the mean and scaled to unit variance. Such a representation of node features has been found efficient in other works~\cite{GNNRE,gnnunlock}.

Next, we describe two node sampling methods that aim to mimic a generic approximation technique.
\vspace{-3mm}
\subsection{Node Sampling}
\label{sec:graph_sampling}

\label{sec:graph:sampling}
As described in Section~\ref{sec:analysis}, the structure of approximate circuits can differ substantially from an exact implementation. However, there are typically fewer nodes and connections in an approximate circuit. Our approach aims to exploit this observation by mimicking a generic approximation constructed from an exact circuit.

In the following, we describe two node sampling methods which remove individual nodes and all transient inputs to this node (the datapath of the node). The number of initially selected nodes (for removal) relates to the level of the approximation. Selecting a large number of initial nodes for removal thus corresponds to a high level approximation, while selecting only a few initial nodes relates to a lower level of approximation. In the following examples, we demonstrate two approaches: (i) \textit{random node sampling} and (ii) \textit{leaf node sampling}. We illustrate these methods on a simple 3-bit adder circuit. Note that 3-bit adder circuits are not included in the actual AppGNN dataset and this is for demonstrative purposes only.

\subsubsection{Random Node Sampling}
\begin{figure}[!t]
 \centering
 \includegraphics[width=0.9\linewidth]{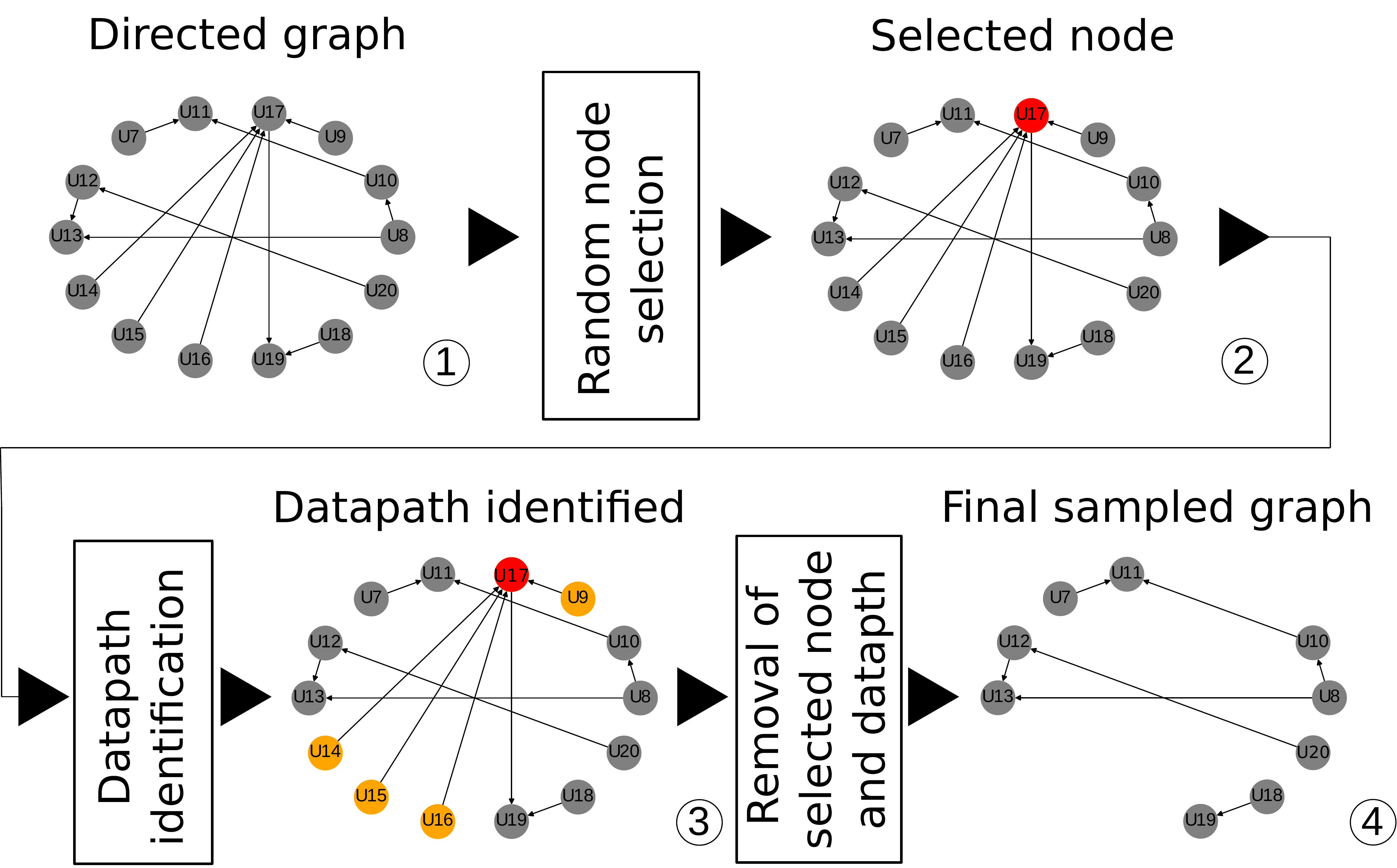}
 \caption{The work flow of our random node sampling.}
 \label{fig:graphsampling:random}
\end{figure}
\label{sec:random:sampling}
We select a number of nodes for removal based on the desired level of approximation. In this example, we are randomly selecting a single node. We then identify the datapath of this node and remove all found nodes (including the initially selected node) from the graph. We now describe this process in detail, which is also displayed in Figure~\ref{fig:graphsampling:random}.

We start with a directed graph that is constructed from a gate-level netlist according to Section~\ref{sec:representation} (Step \bettercircle{1}). Then, we randomly select a node in the graph for removal. U17 (in red) is selected in this example (Step \bettercircle{2}). In order to mimic approximation, we identify the datapath of the selected node using Algorithm~\ref{algo:datapath}. The identified nodes, including the initially selected one, will be marked for removal (Step \bettercircle{3}). Finally, all found nodes are removed from the graph (Step \bettercircle{4}) using Algorithm~\ref{algo:graphsampling}. Depending on the chosen node, the remaining subgraphs can differ substantially. In the example shown in Figure~\ref{fig:graphsampling:random}, node U17 is selected and four additional nodes are identified as the datapath of U17. If node U20 had been chosen instead of U17, only this node would be removed since the datapath is empty as it is a root node.

\subsubsection{Leaf Node Sampling}
\begin{figure}[!t]
 \centering
 \includegraphics[width=0.9\linewidth]{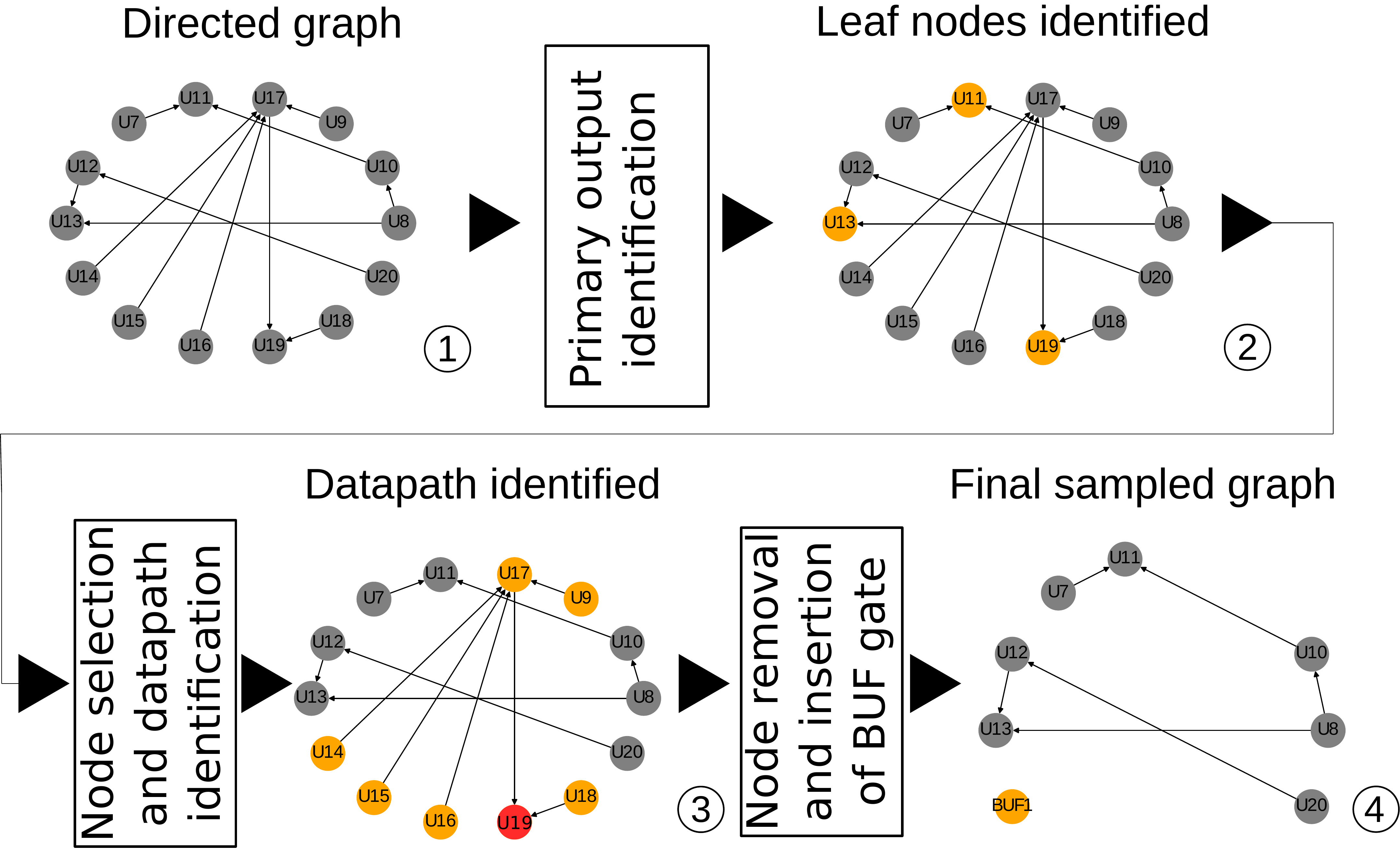}
 \caption{The work flow of our leaf node sampling approach.}
 \label{fig:graphsampling:PO}
\end{figure}
\label{sec:leaf:sampling}
In contrast to the random node sampling technique previously described, we will now explain our leaf node sampling technique. This method is closer to actual approximation since it takes the role of primary outputs (leaf nodes in the graph representation) into account. Depending of the desired level of approximation, this method will select a number of leaf nodes for removal. Leaf nodes in a graph are identified using Algorithm~\ref{algo:leafnodes}. The entire sampling process is displayed in Figure~\ref{fig:graphsampling:PO} and is described in the following.

Similar to the random node sampling technique, we start with a directed graph that is constructed from a gate-level netlist in a way according to Section~\ref{sec:representation} (Step \bettercircle{1}). Next, all leaf nodes that are present in the graph are identified using Algorithm~\ref{algo:leafnodes} (Step \bettercircle{2}). Depending on the desired level of approximation, a number of identified leaf nodes are selected for removal (Step \bettercircle{3}). In this example, only one node (U19, in red) is selected. Thereafter, the datapath(s) of the previously selected leaf node(s) are identified using Algorithm~\ref{algo:datapath} and all found nodes are marked for removal. Finally, all previously marked nodes are removed from the graph (Step \bettercircle{4}) using Algorithm~\ref{algo:graphsampling}. Since leaf nodes (which correspond to primary outputs) have been removed, their datapath is replaced with a single new node. This is to ensure that the resulting graph has the same number of leafs as before and to retain more of the original graph structure. The feature vector $x$ associated with this node will capture that this node is a primary input and a primary output (it redirects all input directly to the output), that the input and output degree are both be one and that the gate type is \texttt{BUF}.

\begin{algorithm}
    \caption{Find all nodes in the datapath of a node}\label{algo:datapath}
    \hspace{-0.6cm} \textbf{Input:} Graph \textit{G(V, E)}; current node $c \in V$; empty list \textit{$N_{DP}$} \\
    \hspace{-3.9cm} \textbf{Output:} a set of nodes $N_{DP} \subseteq V$  
    \begin{algorithmic}[1]
    \Procedure{FindDataPath}{}
    \State $N_{pre} \gets $ predecessors($G$, $c$)
    \If{$c \in N_{pre}$}
    \State $N_{pre}$.remove($c$) \Comment{Avoid loops}
    \EndIf
    \For{$v \in N_{pre}$}                    
        \If{$v \notin N_{DP}$}
            \State FindDataPath($G$, $v$, $N_{DP}$) \Comment{Recursively find nodes}
        \EndIf
    \EndFor
    \If{outputDegree($c$) = 1}
        \State $N_{DP}$.append($c$) \Comment{$c$ is exclusive to this path}
    \EndIf
    \State \textbf{return} $N_{DP}$
    \EndProcedure
    \end{algorithmic}
\end{algorithm}
\vspace{-7.5mm}
\begin{algorithm}
    \caption{Graph sampling}\label{algo:graphsampling}
    \hspace{-2cm} \textbf{Input:} Graph \textit{G(V, E)};  nodes to remove $N_r \subseteq V$ \\
    \hspace{-3.8cm} \textbf{Output:} Sampled graph \textit{G'(V', E')}  
    \begin{algorithmic}[1]
    \Procedure{SampleGraph}{}
    \State $nodesToDrop \gets \{\O\}$
    \For{$i \in N_r$}
        \State $dataPath \gets$ FindDataPath($G$, $i$, $\{\O\}$) 
        \State $nodesToDrop$.append($dataPath$)
    \EndFor
    \State $G' \gets G$.removeNodes($nodesToDrop$)
    \State \textbf{return} $G'$
    \EndProcedure
    \end{algorithmic}
\end{algorithm}
\vspace{-7.5mm}
\begin{algorithm}
    \caption{Identify leaf nodes}\label{algo:leafnodes}
    \hspace{-5.3cm} \textbf{Input:} Graph \textit{G(V, E)}; \\
    \hspace{-4.2cm} \textbf{Output:} A set of nodes $L \subseteq V$   
    \begin{algorithmic}[1]
    \Procedure{IdentifyLeafNodes}{}
    \State $L \gets \{\O\}$
    \For{$v \in V$}
        \If{outputDegree($v$) = $0$}
            \State $L$.append($v$)
        \EndIf
    \EndFor
    \State \textbf{return} $L$
    \EndProcedure
    \end{algorithmic}
\end{algorithm}

\begin{figure*}
 \centering
 \includegraphics[width=\linewidth]{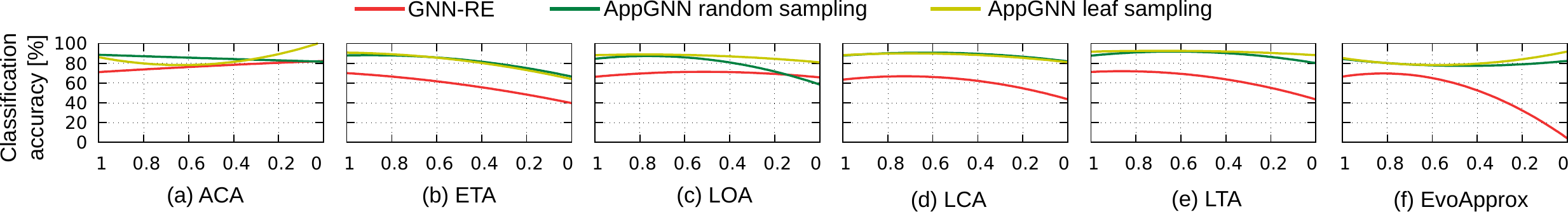}
 \caption{Classification accuracy of GNN-RE~\cite{GNNRE} and AppGNN with random node sampling and with leaf node sampling. The X-axis is the normalized circuit area. A small circuit area correlates with a high approximation aggressiveness, thus the level of approximation increases from left-to-right in each plot.}
 \label{fig:allscores}
\end{figure*}

\vspace{-1.5em}
\subsection{GNN Model}
We employ the fundamental GAT architecture described in Section~\ref{sec:GNN} to perform node classification. We further utilize the node sampling approach, GraphSAINT~\cite{zeng2019graphsaint}, to maintain scalability, as suggested in~\cite{GNNRE}. The GraphSAINT approach samples sub-graphs from the original input graph, and a full GNN is constructed for each extracted sub-graph. We employ two GAT aggregation layers with a hidden dimension of $256$ each and ReLU activation function. For the attention mechanism, $K=8$. The final layer is a fully-connected layer of size $5$ with a Softmax activation function for classification. The GNN is trained using the \textit{Adam optimization algorithm}, with an initial learning rate of $0.01$ and a dropout rate of $0.1$.

\subsection{Dataset Generation}
\label{sec:dataset}
In order to improve the classification accuracy of the GNN on approximate circuits, we extend the training dataset. We do so by adding 35 exact adder circuits of varying bit-widths (see Table~\ref{tab:training:dataset}). After these circuits are converted into their graph representation, as described in Section~\ref{sec:representation}, they are passed through the node sampling stage, as described in Section~\ref{sec:graph:sampling}. During this stage, either random sampling or leaf node sampling is performed. For each bit-width of the exact adders (8,9,12 and 16), 9 sampled graphs (8 for the 8bit adder) with increasing approximation aggressiveness are retrieved. This means that 1 to 9 nodes of a original graph are selected for removal (including their datapath).

An overview over the training dataset used in AppGNN can be found in Table~\ref{tab:training:dataset}. For the training process, this dataset is split as follows: $65\%$ for training, $25\%$ for validation and $15\%$ for testing. 
In the following, we show the evaluation of our AppGNN approach.
\vspace{-0.5em}
\section{Experimental Overview}
We evaluate AppGNN on the AxC dataset shown in Table~\ref{tab:approx:dataset}. We compare the node-level classification accuracy of AppGNN to that of GNN-RE~\cite{GNNRE}, which serves as a baseline.
All datasets consist flat gate-level netlists that are synthesized from their RTL description using Synopsys Design Compiler~\cite{synopsys:designcompiler} using the \texttt{compile\_ultra} directive while performing area optimization. The designs are synthesized using a 14nm FinFET technology library. The conversion from netlist to the graph representation is implemented in \textit{Perl} and \textit{Python3}. Our node sampling methods are also implemented in \textit{Python3}. Training is carried out on a single computer with 16 cores (Intel Core i7-10700 CPU @ 2.90GHz) and 32GB of DDR4 RAM.

We employ the random walk sampler of GraphSAINT with a walk depth of $2$ and $3000$ root nodes. We run training for $100$ epochs. The GNN model is evaluated on the graphs in the validation set after each epoch. The best-performing model on the validation set is restored at the end of training and subsequently used to evaluate the GNN on the testing set.
\vspace{-0.3cm}
\subsection{Results}
\begin{figure}
 \centering
 \includegraphics[width=0.8\linewidth]{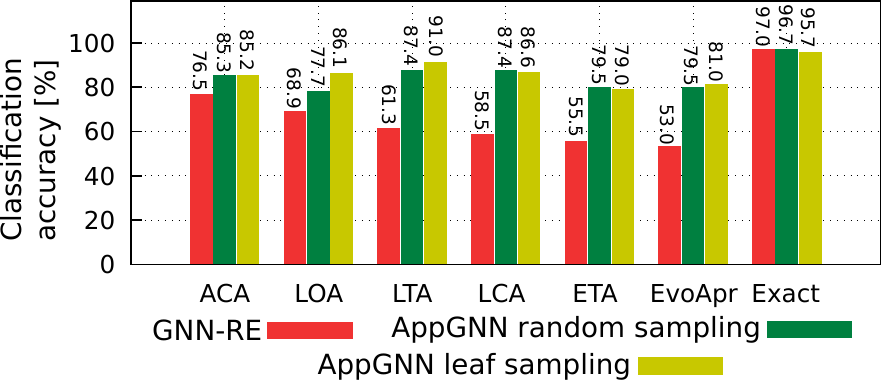}
 \caption{Average classification accuracy of GNN-RE~\cite{GNNRE} and AppGNN with random node sampling and with leaf node sampling for all evaluated classes of adder circuits.}
 \label{fig:accuracy:aprx:adders}
\end{figure}

Our results are presented in the following. Figure~\ref{fig:allscores} shows the classification accuracy of AppGNN and GNN-RE~\cite{GNNRE}, which serves as a baseline, for all classes of AxC adders in Table~\ref{tab:approx:dataset}. In each figure, the X-axis displays the normalized circuit area and thus is related to the level of approximation (which increases from left-to-right on the X-axis). Figure~\ref{fig:accuracy:aprx:adders} summarizes these results and displays the average classification accuracy of GNN-RE and AppGNN. 

\textbf{Comparison to GNN-RE.} As Figure~\ref{fig:allscores} demonstrates, AppGNN outperforms GNN-RE in all benchmarks, regardless of the used sampling method or type of approximation. Except for very high levels of approximation in LOA-based circuits (Figure~\ref{fig:allscores} (c)) when using random node sampling. Here, AppGNN performs circa 8 percentage points worse than GNN-RE. However, AppGNN with leaf node sampling still performs better than GNN-RE in this benchmark. It is noteworthy that, in the case of ACA (Figure\ref{fig:allscores}(a)) and EvoApprox adders (Figure\ref{fig:allscores}(f)) benchmarks, AppGNN with leaf node sampling performs better with an increase of approximation aggressiveness. 

On average, AppGNN outperforms GNN-RE on all types of approximate adder circuits as Figure~\ref{fig:accuracy:aprx:adders} shows. The smallest gains in classification accuracy are achieved on the ACA based circuits with $8.8$ percentage points compared to GNN-RE, which already performs well on this type of circuit. The largest gains can be seen in the EvoApprox adder dataset. Here, AppGNN outperforms GNN-RE by 28 percentage points.

\textbf{Effect of the AxA Type.} With up to $91\%$ classification accuracy, AppGNN performs best on LTA based circuits. This is not a surprising finding; LTA and LCA based circuits are implemented by either truncating $m$ input bits of the operands (LTA) and feeding the output with logical zeroes, or by copying (LCA) $m$ bits of an input operand to the output. In our graph representation, this architecture leads to many isolated nodes (nodes that are both PIs and POs and no adjacent nodes). Our node sampling methods produce graphs that are structurally similar to these, thus allowing AppGNN classify them more accurately. Simultaneously, the classification accuracy of AppGNN on exact adder circuits is comparable to that of GNN-RE, as Figure~\ref{fig:accuracy:aprx:adders} shows. Only a negligible difference of up to $1.3$ percentage points is observed.

\begin{figure}
 \centering
 \includegraphics[width=0.8\linewidth]{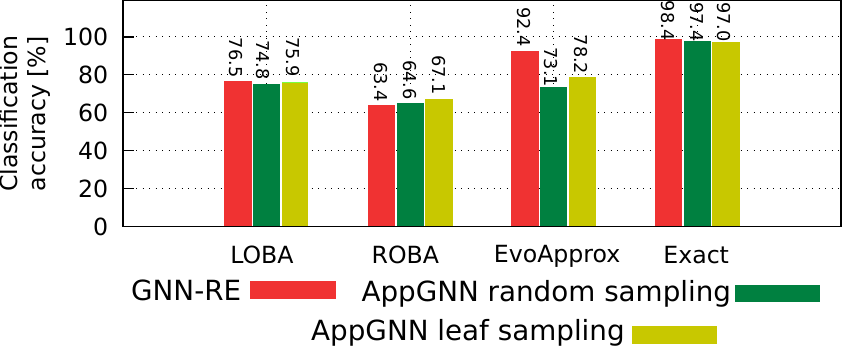}
 \caption{Average classification accuracy of GNN-RE~\cite{GNNRE} and AppGNN with random node sampling and with leaf node sampling for all evaluated classes of multiplier circuits.}
 \label{fig:accuracy:aprx:multipliers}
\end{figure}

\textbf{Effect of the AxM Type.} Figure~\ref{fig:accuracy:aprx:multipliers} illustrates the classification accuracy of GNN-RE and AppGNN on multiplier circuits. In general, GNN-RE does not suffer the same accuracy loss when classifying approximate multipliers as seen in approximate adders. Although AppGNN was not specifically designed to improve the classification accuracy in approximate multipliers, it still managed to outperform GNN-RE by 3.7 percentage points on ROBA based multipliers as Figure~\ref{fig:accuracy:aprx:multipliers} shows. On the LOBA dataset, GNN-RE and AppGNN are on a par with each other, only a negligible difference in classification accuracy can be observed. However, in the case of EvoApprox multipliers, AppGNN suffers significantly. Here, the approximation-unaware GNN-RE actually outperforms AppGNN by 14.2 percentage points. Lastly, the evaluation shows that AppGNN performs similar to GNN-RE when classifying accurate multiplier circuits. A difference of up to 1.4 percentage points can be observed.

\textbf{Effect of the Aggressiveness of the Approximation.} As Figure~\ref{fig:allscores} demonstrates, the classification accuracy of GNN-RE significantly drops with an increase of approximation aggressiveness with the exception of ACA and LOA based circuits. Although a similar trend is observable in the results of AppGNN, it is typically much weaker or even reversed, i.e. in EvoApprox adders (Figure~\ref{fig:allscores}(f)).
\vspace{-0.5em}
\section{Conclusion}
In this work, we investigated the impact of Approximate Computing (AxC) on functional Reverse Engineering (RE). To the best of our knowledge, this is the first time such an investigation is performed.
We demonstrated that traditional means of functional RE are insufficient in the context of AxC. Although Machine Learning (ML)-based methods can handle some variation in the circuit and still provide reasonable results, with an increase of approximation aggressiveness their classification accuracy declines rapidly. We proposed a method for approximation-aware functional RE using Graph Neural Networks (GNNs). Our presented graph sampling based methods aim to mimic the structure of a generic approximation method in order to make a GNN aware of approximation. We evaluated our approach on a wide range of different datasets to show the improved classification accuracy of AppGNN against state-of-the-art GNN-RE. Our extensive evaluation demonstrated how AppGNN outperforms the GNN-RE method in almost all cases.

\vspace{-0.1cm}

\section*{Acknowledgements}
This work was supported in part by the German Research Foundation (DFG) through the Project ‘‘Approximate Computing aCROss the System Stack (ACCROSS)’’ AM 534/3-1,  under Grant 428566201. Besides, this work was also supported by the Center for Cyber Security (CCS) at New York University Abu Dhabi (NYUAD).

\balance
\bibliographystyle{ACM-Reference-Format}
\bibliography{bibliography}
\appendix
\end{document}